\documentclass[runningheads]{llncs}
\usepackage{etoolbox}
\AtBeginEnvironment{thebibliography}{\small \setlength{\itemsep}{0pt} \setlength{\parskip}{0pt}}
\usepackage[T1]{fontenc}
\usepackage{amsmath}
\usepackage{graphicx}
\usepackage{xcolor}
\usepackage{booktabs,tabularx,multirow}
\usepackage{siunitx}
\usepackage{array}        
\usepackage{adjustbox}    

\let\paragraph\relax
\let\subparagraph\relax
\usepackage{titlesec}
\usepackage{url}
\titlespacing*{\section}
  {0pt}{1.2ex plus 0.8ex minus 0.5ex}{0.6ex plus 0.3ex}

\titlespacing*{\subsection}
  {0pt}{0.8ex plus 0.5ex minus 0.3ex}{0.4ex plus 0.2ex}

\begin{document}
\title{KCoEvo: A Knowledge Graph Augmented Framework for Evolutionary Code Generation}
\titlerunning{KCoEvo: KG-Augmented Evolutionary Code Generation}

%
%

\author{
  Jiazhen Kang\inst{1}\inst{3}\thanks{Work done during an internship at ByteDance.}
  \and Yuchen Lu\inst{1}
  \and Chen Jiang\inst{1}
  \and Jinrui Liu\inst{1}
  \and Tianhao Zhang\inst{1}
  \and \\
  Bo Jiang\inst{3}
  \and Ningyuan Sun\inst{3}
  \and Tongtong Wu\inst{2}
  \and Guilin Qi\inst{1}\thanks{Corresponding author. \email{gqi@seu.edu.cn}}
}

\authorrunning{Kang et al.}

\institute{
  Southeast University, Nanjing, China
  \and Monash University, Melbourne, Australia
  \and ByteDance Ltd., China
}

\maketitle

  

%

%
\begin{abstract}
Code evolution is inevitable in modern software development. Changes to third-party APIs frequently break existing code and complicate maintenance, posing practical challenges for developers.
While large language models (LLMs) have shown promise in code generation, they struggle to reason without a structured representation of these evolving relationships, often leading them to produce outdated APIs or invalid outputs.
In this work, we propose a knowledge graph-augmented framework that decomposes the migration task into two synergistic stages: evolution path retrieval and path-informed code generation. 
Our approach constructs static and dynamic API graphs to model intra-version structures and cross-version transitions, enabling structured reasoning over API evolution. 
Both modules are trained with synthetic supervision automatically derived from real-world API diffs, ensuring scalability and minimal human effort.
Extensive experiments across single-package and multi-package benchmarks demonstrate that our framework significantly improves migration accuracy, controllability, and execution success over standard LLM baselines.
The source code and datasets are available at: \url{https://github.com/kangjz1203/KCoEvo}.
\end{abstract}
\section{Introduction}
\label{sect:intro}
Modern software development ecosystems evolve as large-scale, systems driven by third-party libraries and frameworks~\cite{manikas2013software}.
%
Rapid API evolution across versions compromises the compatibility and consistency of dependent projects, creating confusion and maintenance challenges in real-world development.
For example, Microsoft has addressed the complexity of tracking transitively referenced packages in Visual Studio, noting that a typical project may include 20-70 such dependencies when only 6-10 direct dependencies are declared~\cite{MicrosoftTransitiveDependencies2022}.

Large Language Models (LLMs), such as \texttt{LLaMa}~\cite{Touvron2023} and \texttt{GPT-4o}~\cite{gpt4o}, have shown remarkable capabilities in code understanding, bug fixing, and documentation generation~\cite{abs-2203-07814,abs-2107-03374,Bi2025}. 
%
%
%
While LLMs demonstrate impressive capabilities in code synthesis and documentation generation, their reasoning process operates over implicit, unstructured internal knowledge.
This limits their ability to capture temporal and relational dependencies that characterize code evolution.

Empirical studies~\cite{xinp_evo,wang2025codesync,versicode} show that existing LLMs often produce semantically inconsistent or version-incompatible code, primarily due to their inability to model code or API evolution as explicit, queryable knowledge.
These limitations reveal a fundamental gap in current systems: the absence of structured and controllable reasoning mechanisms that can capture version transitions, maintain dependency consistency, and align migration decisions with user intent.
\begin{figure}[!htbp]
    \centering
    \includegraphics[width=\linewidth]{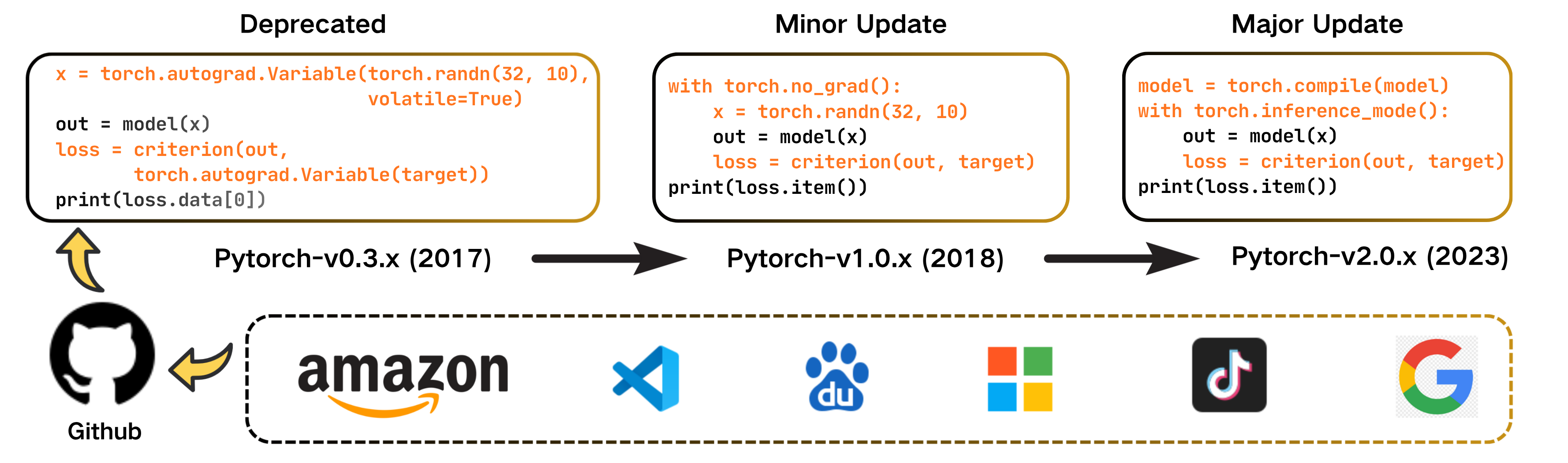}
   \caption{\textbf{Illustration of API evolution in PyTorch across three update stages.}
The figure depicts the progressive modernization of PyTorch’s inference pipeline from \texttt{v0.3.x} (2017) to \texttt{v2.0.x} (2023), covering the transition from \textit{Deprecated} APIs to \textit{Minor} and \textit{Major Updates}.The lower section highlights major \textit{industry adopters} (e.g., Amazon, Microsoft, Baidu, TikTok, Google, and VSCode), with data samples partially collected from their open-source GitHub repositories, emphasizing the practical relevance of version-aware code migration in large-scale production ecosystems.}
    \label{fig:pytorch_evolution}

\end{figure}
Existing retrieval-augmented or prompt-tuning methods provide shallow contextual hints but lack a principled representation of code evolution as structured knowledge.
Consequently, they fail to support symbolic reasoning or graph traversal across versions key operations required for consistent migration.
More critically, these approaches lack a principled way to represent or traverse the structural trajectory of code changes~\cite{jiang2024survey}.
To address this challenge, we propose a knowledge graph-augmented framework that decomposes version-aware code migration into two synergistic subtasks: 
(1) evolution path retrieval, where the model identifies a compatible transition trajectory from the source API call to the target version, and (2) path-informed code generation, where the model generates updated code using the planned path as structural guidance. 
%

The proposed framework adopts a two-level structured knowledge representation, comprising a static API graph that captures intra-version relationships and a dynamic alignment graph that models cross-version transitions~\cite{abs-2402-19173}.
In this representation, API evolution is formalized as a knowledge graph, where each type of evolutionary change is represented as a semantic edge linking related API entities.
These graphs empower the model with a structured deductive capability, compelling its predictions to cohere with both the library's semantic integrity and their evolutionary trajectory. 
%
%
%
To summarize, our main contributions in this paper are: 
\begin{enumerate}
    \item We construct a unified knowledge graph framework that explicitly models API evolution as structured knowledge, capturing both intra-version hierarchies and cross-version transitions. 
    %
    
    \item We formulate version-aware code migration as a two-stage task pipeline, enabling controllable planning and generation.
    
    
    \item We conduct extensive experiments across varied settings, analyzing the effects of graph complexity, reasoning control, and training strategies.
\end{enumerate}

\section{Related Work}
\subsection{Knowledge Graphs for Code Representation and Reasoning}
For a significant period, knowledge graphs have been increasingly adopted in software engineering to represent and reason over structured information related to code.
CodeKG~\cite{abdelaziz2021toolkit}, GraphCodeBERT~\cite{Guo2020GraphCodeBERTPC} and CodeXGraph~\cite{liu2024codexgraph} demonstrate the effectiveness of graph-based encodings in downstream tasks like code summarization, recommendation, and bug detection.
Prior studies have employed various graph-based abstractions to represent code, ranging from syntax-level structures such as abstract syntax trees (ASTs) and control-flow graphs (CFGs) to semantic-level representations such as code knowledge graphs.
%


\subsection{LLM-based Reasoning and Knowledge Integration}
Retrieval-Augmented Generation (RAG)~\cite{lewis2020retrieval,min_RAG} addresses the hallucination in LLMs by grounding their outputs in retrieved data sources. However, it remains limited in maintaining version consistency during code generation, especially when dealing with evolving API semantics~\cite{versicode,oneeval}. To overcome these constraints, recent studies have explored richer integration methods, including retrieval-aware knowledge injection~\cite{chen2024raki}, continual fine-tuning with version-labeled data~\cite{liang2025rustevo,recode}, and graph-based reasoning frameworks that preserve semantic 
structure~\cite{versicode} across evolving entities. 
%
%
%
\subsection{Code Evolution and Migration Tooling}
The continuous evolution of third-party libraries introduces persistent challenges for maintaining code compatibility and developer productivity~\cite{xinp_evo,wu2026environment}. 
More recently, with the rise of large language models, researchers have begun integrating structured API evolution knowledge into model reasoning~\cite{lifespan,wu2026environment}. Empirical evaluations such as \textsc{LLMAPIs}~\cite{xinp_evo} reveal that code LLMs frequently recommend deprecated APIs, reflecting temporal knowledge obsolescence. To address this, retrieval-augmented generation frameworks~\cite{liang2025rustevo} incorporate evolution graphs or version-aware documentation as contextual signals, improving cross-version code generation accuracy. These studies collectively demonstrate that structured evolution knowledge encoded as graphs, rules, or retrievable contexts plays a crucial role in bridging static analysis and LLM-based evolutionary code generation.

\section{Method}
\subsection{Task Definition}
\label{subsec:task}
We define the query $Q$ as the input of LLMs for version-aware code generation.
It consists of a source code snippet $C_{\text{old}}$ that depends on an older API version $V_{\text{old}}$, together with its corresponding functional description $D$.
The objective is to generate a target code snippet $C_{\text{new}}$ that conforms to the updated API version $V_{\text{new}}$  while preserving the original semantic functionality.
Formally, the task can be represented as:
\begin{equation}
    Q = (\, C_{\text{old}},\, V_{\text{old}},\,V_{\text{new}},\, D )
\end{equation}
\begin{equation}
    C_{\text{new}} = f_{\theta}(G_{\text{evo}},\; Q),
\end{equation}
where $f_{\theta}$ denotes LLMs parameterized by $\theta$, 
$G_{\text{evo}}$ represents the structured \textit{API evolution knowledge graph} encoding API entities, relationships, and temporal dependencies across versions, 
and $Q$ denotes the query.

\begin{figure}[!htbp] 
    \centering
    \includegraphics[width=\linewidth]{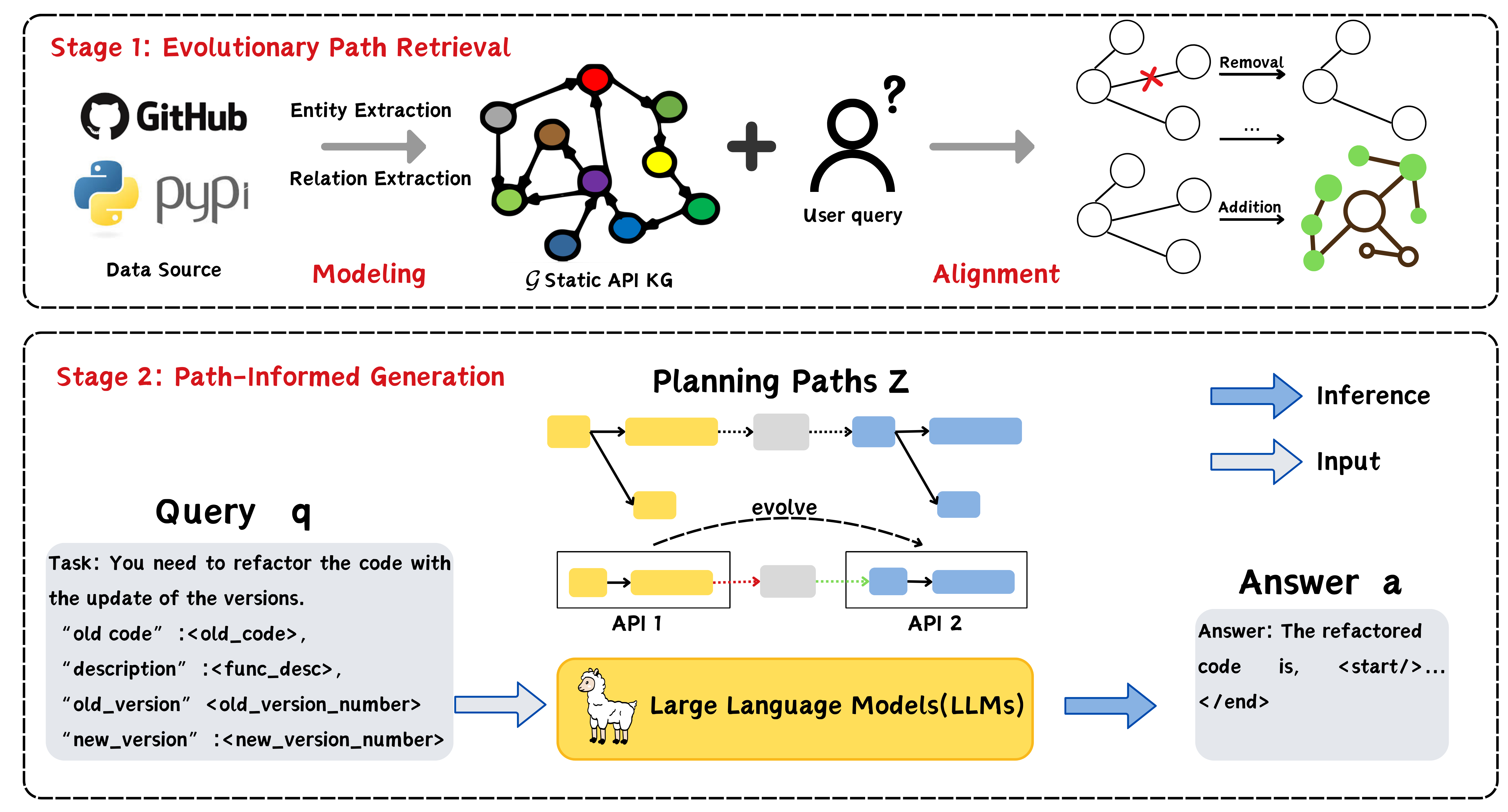}
    \caption{\small Overview of the proposed framework for evolutionary code generation.}
    \label{fig:framework}

\end{figure}

\subsection{Framework Overview}
\label{sect:method}
We represent API evolution through a structured knowledge graph, which serves as a foundation to guide LLMs in generating code aligned with the target version. This pipeline is conceptually divided into two synergistic stages: \texttt{Evolutionary Path Retrieval} and \texttt{Path-Informed Code Generation}.
In the first stage, we first perform \texttt{Graph Modeling} to construct a knowledge graph that encodes API entities, their relationships, and evolutionary patterns. Based on this graph, we then execute \texttt{Dynamic KG Alignment}, where LLMs are employed to identify and rank viable migration trajectories by aligning semantically related API pairs across versions.
In the second stage, \texttt{Path-Informed Code Generation}, the reasoning module leverages these retrieved evolutionary paths to generate version-consistent code that conforms to the target API specifications.

\subsection{Modeling API Evolution with Knowledge Graphs}
\label{sect:graph-construction}
The rapid evolution of APIs, while essential to software progress, undermines LLMs’ ability to generate version-consistent code, thereby introducing syntax and compatibility errors during development. To overcome this challenge, we integrate KGs to explicitly capture both the static structure and dynamic evolution of APIs across versions.
In our framework, each API and its associated evolution types (e.g., rename, relocate, deprecate) are represented as nodes and semantic edges within a version-aware KG.
This structured representation allows LLMs to perform graph-guided reasoning, aligning outdated API calls with their updated counterparts and improving version-aware code migration accuracy.
\begin{figure}[!htbp]
    \centering
    \includegraphics[width=\linewidth]{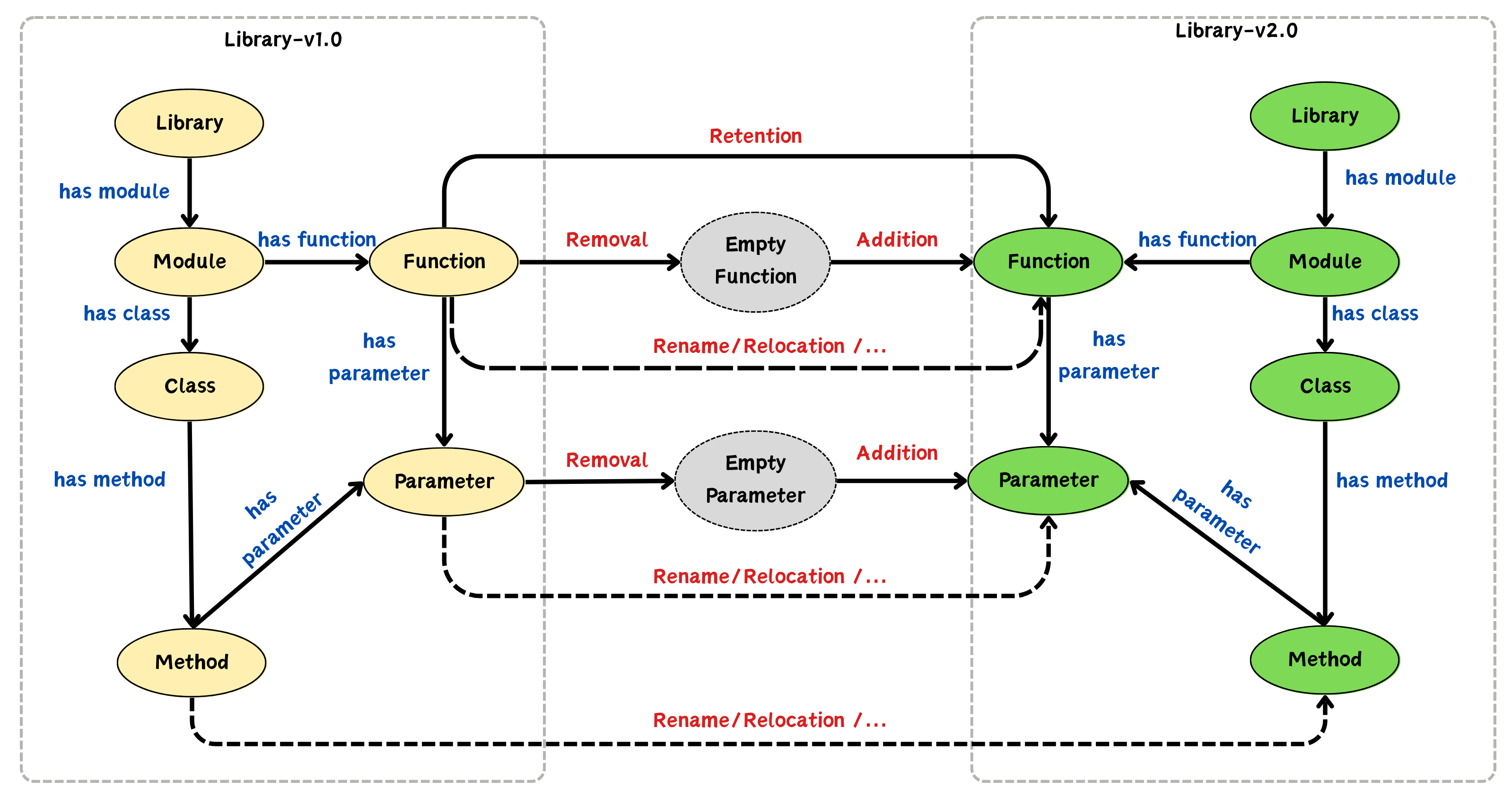}
    \caption{\small{Lightweight schema of the version-aware knowledge graph. Blue edges represent intra-version relations constructed offline; red edges indicate cross-version transitions established dynamically at runtime. 
    Dashed lines, associated with '\texttt{Renaming}' or '\texttt{Relocation}', indicate more complex transformations where an entity evolves through significant modifications to its core attributes or contract. }}
    \label{fig:ontology_schema}
\end{figure}


\subsection{API Knowledge Graph Construction}
The static knowledge graph encodes the structural hierarchy and semantic attributes of APIs within a specific version of a library. It is constructed offline from open-source repositories such as GitHub~\footnote{\url{https://github.com/}} and PyPI~\footnote{\url{https://pypi.org/}} by analyzing the source code and metadata of each release.
Each node in the static graph represents an API entity (e.g., library, module, class, method, function, parameter), and each edge represents a semantic or syntactic relation (e.g., \texttt{has\_function}, \texttt{has\_parameter}, \texttt{returns}, \texttt{has\_description}). The ontology of these types is visualized in Figure~\ref{fig:ontology_schema},  and reflects common structures in Python codebases.
%
For entity extraction, we analyze the hierarchical structure of each library to automatically identify core API elements, including functions, classes, and methods. Using Abstract Syntax Tree (AST) analysis, we extract these entities and capture their key attributes such as parameters, return types, and docstrings.
For relation extraction, we design two complementary modules: one focuses on intra-version relations that capture dependencies and hierarchies within the same API, while the other models inter-version relations to characterize API evolution patterns. The latter employs rule-based matching to detect additions, deletions, and signature modifications across different library versions.
\subsection{Dynamic Graph Alignment across Versions}

While the static knowledge graph models intra-version structural dependencies, effective code migration further requires reasoning over inter-version API changes. 
To this end, we introduce a rule-based dynamic alignment mechanism as shown in Table~\ref{tab:function_evolution_rules}, which identifies semantically related API entities across consecutive library versions in real time.
Given a user query comprising a source code snippet and the target library version, the system first retrieves the relevant subgraph from the static knowledge base corresponding to the source version. 
It then locates the API nodes appearing in the query and traverses the graph using a breadth-first search (BFS) strategy to identify potential evolutionary counterparts. 
%
%
This traversal is guided by explicit version metadata and relation types (\texttt{retain}, \texttt{deprecate}, \texttt{add}), thereby producing an aligned subgraph that encodes valid version transitions.
\begin{table}[!htbp]

\centering
\label{tab-rule-based-patterns}
\caption{Formal definition of function-level evolution relations across versions. 
$F$ and $F'$ denote the sets of functions in the old and new versions, respectively.}
\label{tab:function_evolution_rules}
\small
\setlength{\tabcolsep}{6pt}
\renewcommand{\arraystretch}{1.1}
\begin{tabularx}{\linewidth}{@{}l|l|X@{}}
\toprule
\textbf{Category} & \textbf{Type} & \textbf{Condition} \\
\midrule
\multirow{5}{*}{Function} 
  & Add   
      & $f \notin F \land f \in F'$ \\

  &  Retain (Ret)
      & $f \in F \land f \in F'$ \\

  & Remove (Rem)
      & $f \in F \land f \notin F'$ \\

  & Rename (Ren)
      & $f \in F, f' \in F' \land \text{name}(f) \neq \text{name}(f')$ \\

  & Relocate (Reloc)
      & $f \in F \cap F' \land \text{module}(f) \neq \text{module}(f')$ \\
\bottomrule
\end{tabularx}

\end{table}


\subsection{Knowledge Graph Augment Evolutionary Code Generation}
Our proposed framework enhances the reasoning capability of LLMs through a structured planning mechanism. 
It adopts a two-phase process: first, the model generates explicit reasoning plans that outline API evolution trajectories based on the alignment subgraphs, which is named planning module; second, it executes the reasoning process guided by these planning paths to achieve version-aware code generation.
%
%
Paths are represented as sequences of interconnected entities linked by relations that denote meaningful relationships and interactions within the graph structure.



\subsubsection{Planning Module}
The planning module serves as the core component that bridges the aligned knowledge graph and the final reasoning process. 
In our implementation, this module is instantiated directly as an LLM-based planner that operates on the aligned subgraph obtained from the previous stage. 
Rather than training a separate model, the same large language model is reused for both planning and reasoning phases. 
Given the aligned subgraph by BFS $\mathcal{G}$ and the source code snippet $C$, the LLM planner denoted as $f_\theta$, generates a candidate evolutionary trajectory $z$, which describes how the API in $c$ transitions to its target version. 
Formally, this inference process is expressed as:
\begin{equation}
z =f_{\theta}(C_{old}, \mathcal{G}),
\end{equation}
where the trajectory $z = (\tau_1, \tau_2, \dots, \tau_n)$ is represented as a sequence of evolutionary triplets. 
Complicated evolutionary relations, such as \texttt{relocate}, \texttt{parameter-change}, and \texttt{rename}, are identified during this step to construct precise evolutionary paths across versions.
%
%
For example, if an API $a_1$ was deprecated in an older version while a semantically similar API $a_2$ was introduced in a newer version, emerging with closely aligned parameter definitions and semantic descriptions,
the corresponding segment of the evolutionary path can be represented as $a_1 \xrightarrow{\text{\texttt{relocate}}} a_1' \xrightarrow{\text{\texttt{parameter-change}}} a_2$.
%
\subsubsection{Reasoning Module}
In this phase, following the retrieval of candidate planning paths from the API KG, we augment the LLM generation capabilities by incorporating graph-structured knowledge through graphical representations. The reasoning module, denoted as $f_{\theta'}$, processes a set of paths $Z = \{z_1,z_2,...z_n\} $ obtained from the planning module to generate the target version code $C_{new}$ in accordance with the task instructions $\mathcal{I}$:
\begin{equation}
C_{new} = f_{\theta'}(C_{old}, Z, \mathcal{I})
\end{equation}
During this process, the LLM must accurately implement the reasoning pattern specified within each individual planning path and ensure comprehensive adherence to the complete trajectory.
\section{Experiments}
We formulate the following research questions (RQs) addressing multiple dimensions to systematically evaluate both the effectiveness and generalizability of our proposed approach:
\begin{itemize}
    \item \textbf{RQ1:} Does the proposed framework consistently outperform baseline LLMs in cross-version code migration tasks, demonstrating improvements in both syntactic accuracy and semantic consistency?
    \item \textbf{RQ2:} How do the proposed planning paths contribute to enhancing model interpretability and reasoning effectiveness compared with raw code-based retrieval approaches?
    \item \textbf{RQ3:} To what extent does supervised fine-tuning, particularly when combined with LoRA-based adaptation, improve the backbone LLMs’ capability to generalize across evolving API versions?
\end{itemize}
\subsection{Experiment Settings}
\subsubsection{Models.}
We selected a range of widely recognized open-source and closed-source models to evaluate the performance of our proposed approach. These models encompass prominent families, including GPT \cite{gpt4o}, LLaMA \cite{Touvron2023}, Gemini~\cite{team2023gemini}, Qwen \cite{Bai2023}, and DeepSeek \cite{Bi2025}. For local deployment and inference, we downloaded smaller-scale models (e.g., those with fewer than 20 billion parameters) from Hugging Face\footnote{\url{https://huggingface.co/}}.
\subsubsection{Data Preparation.}
We select \texttt{VersiCode} as our primary benchmark, which is designed to evaluate complementary aspects of model performance. \texttt{VersiCode} focusing on static accuracy in API-level transformations, it is built by systematically mining historical version changes from widely used libraries such as \texttt{TensorFlow}, \texttt{Pandas}, \texttt{Scikit-learn}, \texttt{Transformers}, etc. 
The dataset consists of hundreds of aligned source and target code pairs, each annotated with symbolic evolution paths derived from API knowledge graphs. 
\texttt{VersiCode} supports evaluation along two axes: by evolution type (e.g., rename, relocation, deprecation) and by release timeline (e.g., \texttt{v4.45} to \texttt{v4.46}). 

\subsubsection{Metrics.}
As a static benchmark, \texttt{VersiCode} employs two complementary metrics \texttt{CDC@k} and \texttt{EM@k} to assess model performance.  
First, the \texttt{Critical Diff Check (CDC)}~\cite{versicode} evaluates the functional correctness of API-level transformations based on five rule-based criteria rather than mere textual similarity.  
Specifically, CDC checks:  
(1) whether the generated code includes the core API token;  
(2) whether the code is syntactically valid and can be successfully executed or compiled;  
(3) whether the number of arguments in the generated API call is consistent with the reference implementation;  
(4) whether contextual constructs such as with statements are preserved; and  
(5) whether keyword argument assignments match the reference.  
Second, EM@1 (Exact Match) measures whether the generated API calls match the reference code both semantically and structurally.  
While EM@1 focuses on functional alignment, CDC further penalizes cases that violate execution constraints, such as invalid syntax, incorrect parameter count, or missing contextual tokens.  
Therefore, even in the absence of explicit Pass@k evaluation, a high CDC score reliably indicates that the generated code can compile and execute correctly under the target API version.

\subsubsection{Implementation Details.} To ensure a fair comparison, all 
experiments---including the Base models, RAG baselines, and our 
framework were conducted using identical decoding parameters, specifically, a temperature of $0.0$ and top-$p$ of $1.0$. While the incorporation of 
knowledge graph information inevitably increases the prompt length, we 
ensured that the RAG baselines (Table~\ref{tab:code_block_rag}) were provided with retrieved textual 
contexts of a comparable length. This protocol demonstrates that the 
performance improvements are specifically attributed to the 
\textit{structured evolutionary knowledge} and hierarchical guidance 
encoded in the KG, rather than a mere increase in the input context size.

\begin{table}[!t]
\vspace*{2mm}
\centering
\caption{Performance across migration types with separate CDC@1 and EM@1 columns. 
“Base” denotes the original setting; “+KG” denotes the improved setting under our framework; 
$\Delta$ is the per–metric improvement (+KG $-$ Base). 
The arrow “$\rightarrow$” indicates the direction of migration, where “Major” corresponds to major version changes (e.g., Torch~2.0.0) and “Minor” corresponds to minor version changes (e.g., Torch~2.1.3).}
\label{tab:performance_compact_singlecol_split}
\tiny
\setlength{\tabcolsep}{2.5pt}
\renewcommand{\arraystretch}{1.05}
\begin{tabularx}{\linewidth}{c|c|cc|cc|cc|cc}
\toprule
\multirow{2}{*}{\textbf{Model}} & \multirow{2}{*}{\textbf{Row}} 
& \multicolumn{2}{c|}{\textbf{Major$\rightarrow$Major}} 
& \multicolumn{2}{c|}{\textbf{Major$\rightarrow$Minor}} 
& \multicolumn{2}{c|}{\textbf{Minor$\rightarrow$Major}} 
& \multicolumn{2}{c}{\textbf{Minor$\rightarrow$Minor}} \\
\cmidrule(l){3-10}
& & CDC@1 & EM@1 & CDC@1 & EM@1 & CDC@1 & EM@1 & CDC@1 & EM@1 \\
\midrule
\multirow{3}{*}{DeepSeek-V3} 
 & Base  & 59.52 & 59.52 & 32.83 & 33.84 &  9.26 & 15.74 & 16.67 & 17.74 \\
 & +KG & 96.83 & 100.00 & 79.29 & 94.44 & 75.00 & 95.37 & 53.76 & 93.01 \\
 & $\Delta$ & +37.31 & +40.48 & +46.46 & +60.60 & +65.74 & +79.63 & +37.09 & +75.27 \\
\midrule
\multirow{3}{*}{Qwen2.5-7B-Instruct-Turbo}
 & Base  & 34.13 & 38.89 & 15.15 & 15.66 &  0.93 &  7.41 &  7.53 &  8.60 \\
 & +KG & 61.90 & 83.33 & 40.91 & 68.18 & 29.63 & 41.67 & 17.74 & 39.25 \\
 & $\Delta$ & +27.77 & +44.44 & +25.76 & +52.52 & +28.70 & +34.26 & +10.21 & +30.65 \\
\midrule
\multirow{3}{*}{Qwen2.5-Coder-32B-Instruct}
 & Base  & 38.10 & 38.10 & 16.16 & 16.16 &  7.41 &  7.41 &  6.99 &  8.60 \\
 & +KG & 84.92 & 93.65 & 67.68 & 92.42 & 74.07 & 87.96 & 46.24 & 75.81 \\
 & $\Delta$ & +46.82 & +55.55 & +51.52 & +76.26 & +66.66 & +80.55 & +39.25 & +67.21 \\
\midrule
\multirow{3}{*}{Mistral-Small-24B-Instruct}
 & Base  & 38.10 & 38.10 & 15.66 & 16.16 &  5.56 &  6.48 &  6.45 &  6.45 \\
 & +KG & 72.22 & 91.27 & 38.38 & 67.68 & 36.11 & 67.59 & 33.87 & 68.28 \\
 & $\Delta$ & +34.12 & +53.17 & +22.72 & +51.52 & +30.55 & +61.11 & +27.42 & +61.83 \\
\midrule
\multirow{3}{*}{Llama-3-70B-Instruct-Turbo}
 & Base  & 24.60 & 39.68 & 14.14 & 16.67 &  0.00 & 10.19 &  5.38 &  8.06 \\
 & +KG & 65.87 & 89.68 & 58.59 & 79.80 & 54.63 & 69.44 & 34.95 & 74.73 \\
 & $\Delta$ & +41.27 & +50.00 & +44.45 & +63.13 & +54.63 & +59.25 & +29.57 & +66.67 \\
\midrule
\multirow{3}{*}{Gemini-1.0}
 & Base  & 50.00 & 56.35 & 32.32 & 45.96 &  4.63 & 21.30 & 13.44 & 25.81 \\
 & +KG & 75.40 & 88.89 & 46.46 & 84.34 & 35.19 & 93.52 & 34.95 & 88.71 \\
 & $\Delta$ & +25.40 & +32.54 & +14.14 & +38.38 & +30.56 & +72.22 & +21.51 & +62.90 \\
\midrule
\multirow{3}{*}{Gemini-1.5-Pro-Latest}
 & Base  & 66.67 & 69.05 & 35.35 & 41.41 & 24.07 & 37.04 & 27.96 & 41.40 \\
 & +KG & 95.24 & 100.00 & 92.93 & 98.48 & 72.22 & 98.15 & 66.13 & 90.86 \\
 & $\Delta$ & +28.57 & +30.95 & +57.58 & +57.07 & +48.15 & +61.11 & +38.17 & +49.46 \\
\midrule
\multirow{3}{*}{Gemini-2.0-Pro}
 & Base  & 38.10 & 46.83 & 30.81 & 43.43 &  9.26 & 13.89 & 16.67 & 25.81 \\
 & +KG & 80.95 & 95.24 & 50.51 & 86.87 & 43.52 & 96.30 & 34.95 & 90.32 \\
 & $\Delta$ & +42.85 & +48.41 & +19.70 & +43.44 & +34.26 & +82.41 & +18.28 & +64.51 \\
\midrule
\multirow{3}{*}{GPT-5}
 & Base  & 92.06 & 95.23 & 64.65 & 100.00 & 82.83  & 90.91 & 46.30 & 72.23 \\
 & +KG & 96.83 & 100.00 & 96.97 & 100.00 & 100.00 & 100.00 & 92.60 & 100.00 \\
 & $\Delta$ &  +4.77 &  +4.77  & +32.32  & +0.00 & +17.17  & +9.09 & +46.30 & +27.77 \\
\bottomrule
\end{tabularx}

\end{table}
\subsection{Results and Analysis}
\label{sec:main_results}

\noindent\textbf{Our knowledge-graph-enhanced framework consistently improves both syntactic and semantic accuracy across all migration types.}  
As shown in Table~\ref{tab:performance_compact_singlecol_split}, every model achieves notable performance gains under the +KG setting. 
For instance, \texttt{DeepSeek-V3} improves from $59.52$ to $96.83$ in CDC@1 and from $59.52$ to $100.00$ in EM@1 under the Major$\rightarrow$Major task, yielding a $\Delta$ of $+37.31$ and $+40.48$, respectively. 
Similarly, \texttt{Llama-3-70B-Instruct-Turbo} and \texttt{Qwen2.5-Coder-32B-Instruct} show over $+60$ points gain in EM@1 across Major$\rightarrow$Minor and Minor$\rightarrow$Major migrations, confirming that structured API reasoning substantially boosts model robustness.  

\noindent\textbf{Improvements are especially pronounced in complex cross-version migrations involving subtle semantic shifts.}  
Cross-version transitions such as Major$\rightarrow$Minor and Minor$\rightarrow$Major benefit the most from graph-guided reasoning. 
For example, \texttt{DeepSeek-V3} achieves a $+60.60$ and $+79.63$ improvement in EM@1 on these two types, while \texttt{Mistral-Small-24B} gains $+51.52$ and $+61.11$, respectively. 
These substantial deltas indicate that our framework effectively mitigates semantic drift and preserves API alignment when facing fine-grained version changes.

\noindent\textbf{High-capacity LLMs still require explicit structural grounding for minor-version migrations.}  
Although well-performing models such as \texttt{GPT-5}  demonstrate strong baseline performance (e.g., $>90$ EM@1 on Major$\rightarrow$Major), their improvements remain moderate ($\Delta$ within $+5$ to $+30$). 
This suggests that even state-of-the-art LLMs fail to generalize across small yet semantically critical API updates without structured supervision, underscoring the necessity of explicit knowledge integration for stable code migration.

\begin{table}[!htbp]

\centering
\caption{Performance (EM@1) of Code Block RAG with Multiple Code Sources.
Code sources include downstream application code, library source code, and StackOverflow snippets.}
\label{tab:code_block_rag}
\scriptsize
\setlength{\tabcolsep}{3pt}
\renewcommand{\arraystretch}{1.1}
\begin{tabular}{lcccc}
\hline
\textbf{Model} & \textbf{Major$\rightarrow$Major} & \textbf{Major$\rightarrow$Minor} & \textbf{Minor$\rightarrow$Major} & \textbf{Minor$\rightarrow$Minor} \\
\hline
Qwen2.5-7B-Instruct            & \textbf{38.89} & 15.66 & 7.41  & 8.60 \\
\hspace{1em}+ Downstream Code  & 38.10          & 14.65 & \textbf{15.74} & 7.53 \\
\hspace{1em}+ Library Source   & 38.10          & \textbf{17.17} & 11.11 & \textbf{16.67} \\
\hspace{1em}+ StackOverflow    & 32.30          & 13.03 & 11.11 & 6.99 \\
\hline
Mistral-24B-Instruct           & \textbf{38.10} & 16.16 & 6.48  & 6.45 \\
\hspace{1em}+ Downstream Code  & 35.24          & 16.16 & 7.41  & 8.06 \\
\hspace{1em}+ Library Source   & 38.10          & \textbf{18.18} & \textbf{14.81} & \textbf{13.98} \\
\hspace{1em}+ StackOverflow    & 38.10          & 15.15 & 5.56  & 6.45 \\
\hline
Llama-3-70B-Instruct           & \textbf{40.48} & \textbf{16.67} & 8.33  & 7.53 \\
\hspace{1em}+ Downstream Code  & 38.10          & 16.16 & 6.48  & 7.53 \\
\hspace{1em}+ Library Source   & 38.10          & 12.42 & \textbf{10.56} & \textbf{14.52} \\
\hspace{1em}+ StackOverflow    & 38.89          & 16.16 & 6.48  & 7.53 \\
\hline
\end{tabular}

\end{table}
\noindent\textbf{Retrieval context significantly influences cross-version generalization.}  
To further investigate \texttt{RQ2}, we evaluate similarity-based retrieval augmentation under three external code sources: downstream application code, library source code, and Stack Overflow snippets. 
As summarized in Table~\ref{tab:code_block_rag}, using library source code as the retrieval context yields the highest improvements, particularly in minor$\rightarrow$minor and minor$\rightarrow$major migrations. 
In contrast, Stack Overflow snippets bring limited gains due to their fragmented and context-poor nature, reinforcing the importance of structurally aligned retrieval contexts for reasoning consistency. The consistent performance gains across diverse models and migration scenarios confirm that our framework delivers robust version-awareness and interpretable reasoning for evolutionary code generation. 
In summary, these results provide empirical support for both \texttt{RQ1} and \texttt{RQ2}, demonstrating that structured knowledge grounding not only enhances code correctness but also improves interpretability under version-evolving settings.

\begin{table}[!htbp]
\centering
\caption{Performance (EM@1) of LoRA fine-tuning across version transition types.
\textit{Planning Paths} provide graph-guided supervision based on API evolution knowledge. 
Their combination significantly improves cross-version code generation performance.}
\label{tab:ablation}
\scriptsize
\setlength{\tabcolsep}{2.5pt}
\renewcommand{\arraystretch}{1.1}
\begin{tabular}{lcccc}
\hline
\textbf{Model \& Configuration} & \textbf{Maj$\rightarrow$Maj} & \textbf{Maj$\rightarrow$Min} & \textbf{Min$\rightarrow$Maj} & \textbf{Min$\rightarrow$Min} \\
\hline
Qwen2.5-Coder-7B-Instruct       & 23.81 & 8.08 & 5.56 & 4.30 \\
\hspace{1em}+ Planning Paths           & \textbf{66.67} & 46.46 & 38.89 & 37.10 \\
\hspace{1em}+ LoRA                     & 27.08 & 7.78 & 8.97 & 4.49 \\
\hspace{1em}+ LoRA + Planning Paths    & 55.56 & \textbf{48.48} & \textbf{57.41} & \textbf{43.01} \\
\hline
Meta-Llama-3-8B-Instruct         & 25.40 & 10.10 & 6.48 & 4.84 \\
\hspace{1em}+ Planning Paths           & \textbf{89.68} & \textbf{85.35} & \textbf{86.11} & 88.17 \\
\hspace{1em}+ LoRA                     & 23.81 & 11.11 & 8.33 & 3.23 \\
\hspace{1em}+ LoRA + Planning Paths    & 88.89 & 84.85 & 86.11 & \textbf{82.26} \\
\hline
DeepSeek-Coder-7B-Base-v1.5      & 3.17 & 0.51 & 1.85 & 1.08 \\
\hspace{1em}+ Planning Paths           & 26.98 & \textbf{33.84} & \textbf{27.78} & 22.04 \\
\hspace{1em}+ LoRA                     & 16.67 & 3.03 & 2.78 & 0.54 \\
\hspace{1em}+ LoRA + Planning Paths    & \textbf{39.68} & 22.22 & \textbf{27.78} & \textbf{23.66} \\
\hline
\end{tabular}

\end{table}

\subsection{Ablation study}

To investigate \texttt{RQ3} , we conduct an ablation study comparing models trained with and without LoRA-based adaptation, as well as configurations that either include or omit the use of planning paths. 
As shown in Table ~\ref{tab:ablation}, incorporating supervised fine-tuning guided by evolutionary path information significantly enhances the model’s ability to handle cross-version code generation. 
The analysis reveals that combining lightweight parameter tuning with structured evolutionary supervision bridges the gap between general-purpose LLMs and version-aware reasoning.
\subsection{Case Studies and Error Analysis}
\begin{figure}[!htbp]

\centering
\includegraphics[width=\linewidth]{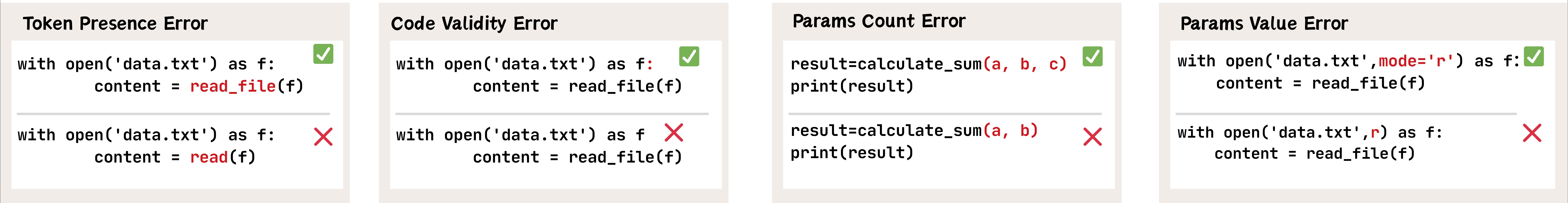}
\caption{
Representative error types observed in version-aware code migration.
The examples illustrate typical reasoning failures encountered by LLMs, including (1) \texttt{Token Presence Error}, (2) \texttt{Code Validity Error}, (3) \texttt{Parameter Count Error} and (4) \texttt{Parameter Value Error}.
These cases highlight the limitations of surface-level pattern learning and motivate the need for structured, knowledge-driven reasoning in API evolution tasks.
}
\label{fig:error_cases}

\end{figure}

Although integrating the knowledge graph substantially improves models' awareness of API evolution, it also introduces new failure patterns that affect code quality. 
Our qualitative analysis (Figure~\ref{fig:error_cases}) focuses on model outputs generated under the \texttt{+KG} setting, which still exhibit compilation failures and logical inconsistencies despite higher semantic alignment scores.

As shown in Table~\ref{tab:performance_compact_singlecol_split}, models such as \texttt{Gemini-1.0}, \texttt{Mistral-Small-24B}, and \texttt{Qwen2.5-Coder-32B} achieve large EM@1 gains ($\Delta$EM@1: +60--80\%), yet their CDC@1 improvement remains considerably smaller. 
This divergence indicates that while KG-enhanced reasoning helps models memorize and match API usage semantically, it does not guarantee syntactically valid or executable code.
In particular, we frequently observe \texttt{Code Validity Error} and \texttt{Parameter Count Error}, where the generated code violates argument signatures or includes missing default parameters,  typical causes of compilation failure.
These findings suggest that explicit evolutionary knowledge can overfit models toward symbolic matching rather than structural consistency.
In practice, this results in code that appears semantically plausible (high EM@1) but fails under strict syntactic or runtime constraints (low CDC@1). 
We hypothesize that integrating  lightweight policy gradient in code generation tasks with compilation feedback could mitigate such inconsistencies, bridging the semantic-syntactic gap.
\subsection{Discussion}
This study demonstrates that integrating structured code evolution knowledge into large language models can substantially enhance both the reliability and interpretability of evolutionary code generation. 
However, continuously updating and storing large-scale knowledge graph repositories across multiple software versions incurs notable computational and energy overhead, raising concerns about sustainability and cost efficiency in large organisations. 
Future efforts should therefore focus on developing lightweight, incremental graph update strategies and adaptive integration mechanisms that can seamlessly align evolving code knowledge with LLM reasoning at scale.

\section{Conclusion and Future Work}
In this paper, we construct a knowledge graph to capture the structural dynamics of API evolution and integrate it into large language models, allowing them to reason over version-specific changes and generate evolution-consistent code across software versions.
Our work presents a knowledge graph-retrieval-based framework that bridges the gap between the reasoning capabilities of LLMs and the practical versioning constraints of modern software development. 
The experimental results reveal that current LLMs struggle to remain synchronized with rapidly evolving APIs, primarily due to temporal knowledge obsolescence. 
In future work, we plan to extend this framework toward broader code intelligence scenarios, such as cross-lingual adaptation, cross-framework compatibility, and developer behavior analysis, beyond the current version-specific scope. 
In parallel, we plan to extend our framework as a plugin integrated into integrated development environments, offering developers proactive, version-consistent code suggestions and automated dependency updates.

%

\section{Acknowledgement}
This research was supported by National Social Science Foundation Key Program of China (No.23ZD222) and ByteDance Inc. We thank the Big Data Computing Center of Southeast University for providing the facility support on the numerical calculations in this paper.
%
%
%
\bibliographystyle{splncs04}
\bibliography{reference_revised}
\end{document}